\newcommand{\bee}{\begin{equation}}
\newcommand{\ene}{\end{equation}}
\newcommand{\beea}{\begin{eqnarray}}
\newcommand{\enea}{\end{eqnarray}}

\documentclass[aps,preprint,pre,showpacs,superscriptaddress]{revtex4}
\usepackage{amsmath} 
\usepackage{graphicx}
\usepackage{dcolumn}
\usepackage{bm}
\begin{document}
\title{Wavefunction of Plasmon Excitations with Space Charge Effects}
\author{M. Akbari-Moghanjoughi}
\affiliation{Faculty of Sciences, Department of Physics, Azarbaijan Shahid Madani University, 51745-406 Tabriz, Iran}

\begin{abstract}
The one dimensional (1D) driven quantum coupled pseudoforce system governing the dynamics of collective Langmuir electron oscillations is used in order to investigate the effects of variety of space charge distributions on plasmon excitations of a nearly free electron gas with arbitrary degree of degeneracy and electron fluid temperature. A generalized closed form analytic expression for the grand wavefunction of collective excitations in presence of an arbitrary space charge distribution is presented based on the stationary solutions of the driven coupled pseudoforce system which has been derived from the Schr\"{o}dinger-Poisson model. The wavefunction and electrostatic potential profiles for some especial cases such as the Heaviside charge distribution, Dirac charge sheet, impurity charge sheet in 1D plasmonic lattice and the Kroning-Penney Dirac charge distributions with particular applications in plasmonics and condensed matter physics is investigated in this paper. It is remarkably found that two parallel Dirac charged sheets completely shield all interior plasmon excitations with any given energy value from outside electrostatic fields and charge densities. It is also found that the presence of even a weakly charged impurity layer within a perfect 1D plasmonic crystal profoundly destroys the periodic electrostatic field of the crystal lattice, hence, the Bloch character of the wavefunction considered in band gap theory of solids. Current investigation of electron excitations in arbitrary degenerate electron gas in the presence of static charge distributions may be used to develop analytic models for a variety of real physical situations. It also helps in further developments of the rapidly growing fields of nanotechnology and plasmonics.
\end{abstract}

\date{\today}

\maketitle

\section{Introduction}

Collective effects play a fundamental role in many body fermionic and bosonic systems like solids \cite{kittel,ash}. In fermionic systems like plasmas multi-ingredient aspects lead to variety of interesting linear and nonlinear properties due to colorful interactions among different species \cite{chen,krall,drazin,klimontovich}. Therefore, a lot of analytical as well as numerical methods have been developed over the years to investigate these interesting aspects of plasmas \cite{rowlands,mitsuo,davidson,sagdeev,karpman}. Almost all physical properties of condensed matter, biological and molecular systems, electron-hole dynamics in solids etc. are affected by complex collective electromagnetic interactions \cite{markovich,man,haug,gardner,ae}. Dynamic properties of such complex settings are usually dealt with kinetic and fluid models \cite{manfredi,shukla1,shukla2,haasbook}. In dense quantum systems with extra quantum potential interactions, on the other hand, the statistical tools to study the field-density time evolution require a memory-time consuming large scale numerical simulations through well-known models like Hartree-Fock, Wigner-Poisson-Maxwell, magnetohydrodynamic and ground state density functional, etc. due to large degree of freedom and complex nature of particle interactions. Nowadays, however, very realistic simulations has been possible for kiloscale degree of freedom (DoF) systems with the help of improved computing systems and algorithms which brings into view some interesting statistical behavior of many body charged systems \cite{matt}. Recently, the generation of high-harmonic in metallic nanoparticles has been examined by numerical evaluation of nonlinear electron dynamics using the quantum hydrodynamics model incorporating the self-consistent Hartree potential, in which collective excitations in 200 gold nanoparticles has been examined \cite{hurst1}. In another work the extended quantum hydrodynamic model with Hartree potential and exchange effect has been employed to numerically study the plasmon excitations in a quantum trap \cite{hurst2}.

Plasmons are elementary collective excitations of free electron systems such as plasmas, metals, nano-metallic compounds, etc. which cause peculiar quantum features in physical properties like electric and thermal conductivities, optical properties and dielectric responses etc. \cite{maier}. Electron-hole quantum transport and tunneling through potential barriers in integrated circuits are typical collective phenomenon understanding of which requires in depth knowledge of these elementary excitations \cite{chenming,seeger,bart}. Study of plasmon excitations has started with pioneering works of Bohm \cite{bohm}, Pines \cite{pines} and Levine \cite{levine}. There has been however a renewed momentum in investigation of different aspects of these entities due to their fundamental contribution to the rapidly growing fields of nonotechnology, optoelectronics, plasmonics and low dimensional systems. For instance, there has been numerous recent investigations revealing many outstanding new quantum aspects of electron plasma oscillations in collective charged environments using density functional, quantum hydrodynamics and kinetic approaches \cite{ydj3,ydj4,dub3,dub4,garcia,marklund,stenflo,shukla0,shuklaprl,nshukla,bret,se,akb1,mend,es,brod,sten,stenf}. In this paper we use the newly introduced concept of coupled pseudoforce \cite{akbpseudo1,akbpseudo2} based on the Schr\"{o}dinger-Poisson to obtain the grand wavefunction which includes the wavefunction together with the collective electrostatic potential of the free electron gas with arbitrary degree of electron degeneracy and electron fluid temperature in Sec. II. This approach is then extended to include the effect of static charge distribution as the driving pseudoforce in Sec. III. The later approach is then used to study the plasmon excitations in Heaviside in Sec. IV and Dirac delta charge distributions and Kronig-Penney model in Sec. V. Conclusions are also presented in Sec. VI.

\section{Grand Wavefunction of Plasmon Excitations}

We would like to study plasmon excitations of finite temperature arbitrary degenerate electron gas in presence of different electrostatic potential configurations solving for the plasmon wavefunction and the electrostatic potential function. Let us now consider a one dimensional isothermal electron gas with the chemical potential $\mu_0$ and fluid temperature $T$. Dynamics of electron fluid is described using the following effective Schr\"{o}dinger-Poisson model which is related to the hydrodynamic formulation as described elsewhere \cite{manfrediprb}
\begin{subequations}\label{sp}
\begin{align}
&i\hbar \frac{{\partial \cal N }}{{\partial t}} =  - \frac{{{\hbar ^2}}}{{2m}}\frac{{\partial {{\cal N} ^2}}}{{\partial {x^2}}} - (e\phi-\mu){\cal N},\\
&\frac{{\partial {\phi ^2}}}{{\partial {x^2}}} = 4\pi e n,
\end{align}
\end{subequations}
in which ${\cal N} =\sqrt{n(x,t)}\exp[iS(x,t)]$ is the time dependent electron fluid wavefunction with ${\cal N}{\cal N^*}=n(x,t)$ and $n(x,t)$ and $u(x,t)=(1/m)\partial S(x,t)/\partial x$ being the number density and fluid speed of the arbitrary degenerate electron gas, respectively. The isothermal electron gas is described via the following generalized equation of state
\begin{subequations}\label{np}
\begin{align}
&{n(\mu,T)} = \frac{{{2^{1/2}}m{^{3/2}}}}{{{\pi ^2}{\hbar ^3}}}  \int_{0}^{ + \infty } {\frac{{\sqrt{{\epsilon}} d{\epsilon}}}{{{e^{\beta ({\epsilon}-\mu)}} + 1}}},\\
&{P(\mu,T)} = \frac{{{2^{3/2}} m{^{3/2}}}}{{3{\pi ^2}{\hbar ^3}}}\int_0^{ + \infty } {\frac{{{{\epsilon}^{3/2}} d{\epsilon}}}{{{e^{\beta ({\epsilon} - {\mu})}} + 1}}.}
\end{align}
\end{subequations}
satisfying the thermodynamic identity $\partial P(\mu,T)/\partial \mu = n(\mu,T)$ which has been used to replace the electron statistical pressure with the chemical potential in the Schr\"{o}dinger-Poisson model. Each electron in the system is described by a wavefucntion which is in the linear perturbation limit a combination of single particle wavefucntion and a collective electrostatic interaction wavefunction as described elsewhere \cite{akbnew}. The later is attributed via the mean field Hartree potential in the Schr\"{o}dinger-Poisson model \cite{hurst2}. We consider the later model as described in terms of the electron chemical potential instead of the scalar Fermi pressure which is assumed to remain constant to the linear order. Therefore assuming ${\cal N} =\psi_1(x)\psi_2(t)$ the coupled pseudoforce model describing the stationary plasmon solution is as follows
\begin{subequations}\label{pf}
\begin{align}
&\frac{{d^2{\Psi(x)}}}{{d{x^2}}} + \Phi(x) = - 2 E \Psi(x),\\
&\frac{{d^2{\Phi(x)}}}{{d{x^2}}} - {\Psi}(x) = 0,
\end{align}
\end{subequations}
in which assuming $\mu\equiv \mu_0$, we have used the normalized quantities in (\ref{sp}) as $\Psi(x)=\psi_1(x)/\sqrt{n_0}$ where $n_0$ is the equilibrium number density of the electron fluid, $\Phi(x)=e\phi(x)$, $E=(\epsilon-\mu_0)/2E_p$ ($E_p=\hbar\omega_p$ with $\omega_p=\sqrt{4\pi e^2 n_0/m}$ being the quantized plasmon energy) and $x=x/\lambda_p$ with $\lambda_p=2\pi/k_p$ and $k_p=\sqrt{2mE_p}/\hbar$ being the characteristic plasmon wavelength. Also, the time dependent part of the wavefunction reads $\psi_2(t)=\exp(-i\omega t)$ with $\epsilon=\hbar\omega$ being the energy eigenvalues.

The grand wavefunction ${\cal G}(x,t)$ characterizing the pseudoforce system \ref{pf}(a) and \ref{pf}(b) with the boundary conditions $\Phi'(0)=\Psi'(0)=0$, $\Phi(0)=\Phi_0$ and $\Psi(0)=\Psi_0$ is
\begin{equation}\label{wf}
{\cal G}(x,t) =\left[ {\begin{array}{*{20}{c}}
{\Phi (x)}\\
{\Psi (x)}
\end{array}} \right] e^{-i\omega t} = \frac{e^{-i\omega t}}{{2\alpha }}\left[ {\begin{array}{*{20}{c}}
{{\Psi _0} + k_2^2{\Phi _0}}&{ - \left( {{\Psi _0} + k_1^2{\Phi _0}} \right)}\\
{ - \left( {{\Phi _0} + k_1^2{\Psi _0}} \right)}&{{\Phi _0} + k_2^2{\Psi _0}}
\end{array}} \right]\left( {\begin{array}{*{20}{c}}
{{{\rm{e}}^{i{k_1}x}}}\\
{{{\rm{e}}^{i{k_2}x}}}
\end{array}} \right),
\end{equation}
in which $E=\hbar\omega$ is the energy eigenvalue and
\begin{equation}\label{eks}
{k_1} = \sqrt {E - \alpha },\hspace{3mm}{k_2} = \sqrt {E + \alpha },\hspace{3mm}\alpha  = \sqrt {{E^2} - 1}.
\end{equation}
Note that $k_1=1/k_2$ holds for any energy eigenvalue $E>1$.

\section{Wavefunction of Driven Plasmon}

The following driven pseudoforce system \cite{akbpseudo1} describes the plasmon excitations in the presence of given static charge distribution due to ions or impurity charge in crystals
\begin{subequations}\label{dpf}
\begin{align}
&\frac{{d^2{\Psi(x)}}}{{d{x^2}}} + \Phi(x) = - 2 E \Psi(x),\\
&\frac{{d^2{\Phi(x)}}}{{d{x^2}}} - {\Psi}(x) = n_i(x).
\end{align}
\end{subequations}
For the sinusoidal space charge distribution $n_i(x)=C\exp(iK x)$ (with $C$ being the normalized charge density) the particular solution to time -independent system (\ref{pf}) is
\begin{subequations}\label{sol1}
\begin{align}
&\Phi_p (x) = \frac{C}{{2\alpha }}\left[ {\frac{{\left( {1 + \beta k_2^2} \right){e^{i{k_1}x}} - \left( {1 + \beta k_1^2} \right){e^{i{k_2}x}} - 2\alpha \beta {e^{iKx}}}}{{\left( {{K^2} - k_1^2} \right)\left( {{K^2} - k_2^2} \right)}}} \right],\\
&\Psi_p (x) = \frac{C}{{2\alpha }}\left[ {\frac{{\left( {{K^2} - k_1^2} \right){e^{i{k_2}x}} - \left( {{K^2} - k_2^2} \right){e^{i{k_1}x}} - 2\alpha {e^{iKx}}}}{{\left( {{K^2} - k_1^2} \right)\left( {{K^2} - k_2^2} \right)}}} \right].
\end{align}
\end{subequations}
where $\beta  = {K^2} - 2E$. Hence, the solution to system (\ref{dpf}) is the combination of general (\ref{wf}) and particular (\ref{sol1}) solutions. It is however easy to give a generalized particular solution to the case of arbitrary driven pseudoforce $n_i(x)=\sum {{C_m}\exp } \left( {iK_mx} \right)$ with $K_m=mK$ as follows
\begin{subequations}\label{solg}
\begin{align}
&\Phi_p (x) = \sum\limits_{m =  - \infty }^{ + \infty } {\frac{{{C_m}}}{{2\alpha }}\left[ {\frac{{\left( {1 + \beta k_2^2} \right){e^{i{k_1}x}} - \left( {1 + \beta k_1^2} \right){e^{i{k_2}x}} - 2\alpha \beta {e^{iK_mx}}}}{{\left( {{K_m^2} - k_1^2} \right)\left( {{K_m^2} - k_2^2} \right)}}} \right]},\\
&\Psi_p (x) = \sum\limits_{m =  - \infty }^{ + \infty } {\frac{{{C_m}}}{{2\alpha }}\left[ {\frac{{\left( {{K_m^2} - k_1^2} \right){e^{i{k_2}x}} - \left( {{K_m^2} - k_2^2} \right){e^{i{k_1}x}} - 2\alpha {e^{iK_mx}}}}{{\left( {{K_m^2} - k_1^2} \right)\left( {{K_m^2} - k_2^2} \right)}}} \right]}.
\end{align}
\end{subequations}
where $C_m=(K/2\pi)\int {n_i(x)\exp ( - iK_m x)dx}$ are the Fourier components of the charge distribution. Note that (\ref{solg}) is the most general analytic particular solution to the plasmon wavefunction in the presence of an arbitrary one dimensional charge distribution $n_i(x)$.

\section{Heaviside Charge Distribution}

There are obviously physical instances in which an ambient free electron gas arrives at a uniform step charge density. However, it is valuable to derive expressions for probability charge density and electrostatic field in the two regions for given energy of the plasmon expiation. Assuming that the for there is an abrupt increase in the ion charge density in the region $x>0$, we have $n_i(x)=A\Theta(x)$ in which $\Theta(x)$ is the Heaviside step function. Note that the current case with charge density barrier is a new aspect of the pseudoforce model of plasmon excitations which is fundamentally different from the case involving the potential barrier considered in many classical quantum problems of Schr\"{o}dinger equation. It can be easily confirmed that the following particular solutions satisfy the driven pseudoforce model with the Heaviside step density distribution, $n_i(x)=A\Theta(x)$.
\begin{subequations}\label{sols}
\begin{align}
&{\Phi _p}(x) = \frac{{A\Theta (x)\left[ {4\alpha E - {\gamma _1}\cos \left( {{k_2}x} \right) + {\gamma _2}\cos \left( {{k_1}x} \right)} \right]}}{{2\alpha }},\\
&{\Psi _p}(x) =  - \frac{{A\Theta (x)\left[ {1 + {\gamma _2} - \cos \left( {{k_2}x} \right) - {\gamma _2}\cos \left( {{k_1}x} \right)} \right],}}{{1 + {\gamma _2}}}.
\end{align}
\end{subequations}
where $\gamma_1=1-2E k_1^2$ and $\gamma_2=1-2E k_2^2$.

Figure 1 shows the numerical evaluation of the solution (\ref{sols}) for different values of the charge density values. The initial values $\Phi_0=\Psi_0=1$ is chosen for all simulations in this paper. However, the continuity of the solutions and the derivative of $\Psi$ is found to be completely independent of the initial values. Figure 1(a) shows the electrostatic potential variations of plasmon excitation in the region of space charge distribution for two different values of negative (thin curves) and positive (thick curves) charge distributions and for a given value of the plasmon energy. It is remarked that the average electrostatic potential level of plasmon for positive/negative space charge distribution is positive/negative. Moreover Fig. 1(b) shows the electrostatic potential variations for an elevated plasmon energy indicating that, while the spacial fluctuation frequency increases due to increase in plasmon energy in both $x<0$ and $x>0$ regions for positive value of $A=0.5$, these fluctuations for negative value of space charge distribution $A=-0.5$ decreases. On the other hand, Figs. 1(c) and 1(d) show the variations of the wavefunction of electrons for the values used in Figs. 1(a) and 1(b). It is revealed that, the average fluctuation level of the wavefunction in Fig. 1(c) resides in negative/positive values for positive/negative space charge distributions in the region $x>0$. This is opposite to the case for electrostatic potential profiles in Fig. 1(a). The same feature is also seen for Fig. 1(d) which depicts the wavefunction variations for an elevated plasmon energy. Another feature revealed by Fig. 1(d) is that increase in the plasmon energy tends to further separate the average fluctuation levels corresponding to negative and positive space charge distribution values, while increasing the fluctuation frequency for both negative and positive space charge distribution signs.

\section{Kronig-Penney Plasmonic Model}

The case of Dirac delta charge distribution, $n_i(x)=A\delta(x)$, may be realized as a coated uniform charge sheet sandwiched in a Josephson junction. The particular solution of driven coupled pseudoforce system may be written as follows
\begin{subequations}\label{sold}
\begin{align}
&{\Phi _p}(x) = \frac{{A\left[ {\Theta (x) - \Theta (0)} \right]}}{{2\alpha }}\left[ {k_2^3\sin \left( {{k_1}x} \right) - k_1^3\sin \left( {{k_2}x} \right)} \right],\\
&{\Psi _p}(x) = \frac{{A\left[ {\Theta (x) - \Theta (0)} \right]}}{{2\alpha }}\left[ {{k_1}\sin \left( {{k_2}x} \right) - {k_2}\sin \left( {{k_1}x} \right)} \right],
\end{align}
\end{subequations}
in which $\Theta(x)$ is the Heaviside theta function assuming $\Theta(0)=1/2$. The solution (\ref{sold}) may be generalized to the case with one dimensional plasmonic crystal with $n_c(x)=B\cos(G x)$ ($G=2\pi/a$) \cite{}. The initial values $\Phi_0$ and $\Psi_0$ in this case are chosen so that the general solution satisfies the lattice periodicity condition. The complete, i.e., general+particular solution for plasmonic crystal with a single Dirac delta charge at the origin is given below
\begin{subequations}\label{soldp}
\begin{align}
&\Phi (x) =  - \frac{{B\left( {{G^2} - k_1^2 - k_2^2} \right)\cos \left( {Gx} \right)}}{{\left( {{G^2} - k_1^2} \right)\left( {{G^2} - k_2^2} \right)}} + \frac{{A\left[ {\Theta (x) - \Theta (0)} \right]}}{{2\alpha }}\left[ {k_2^3\sin \left( {{k_1}x} \right) - k_1^3\sin \left( {{k_2}x} \right)} \right],\\
&\Psi (x) =  - \frac{B\cos \left( {Gx} \right)}{{\left( {{G^2} - k_1^2} \right)\left( {{G^2} - k_2^2} \right)}} + \frac{{A\left[ {\Theta (x) - \Theta (0)} \right]}}{{2\alpha }}\left[ {{k_1}\sin \left( {{k_2}x} \right) - {k_2}\sin \left( {{k_1}x} \right)} \right],
\end{align}
\end{subequations}

Moreover, for a symmetric sheets located at $x=\pm b$ (the parallel charged sheet configuration) the solution for $b>0$ reads
\begin{subequations}\label{sold2}
\begin{align}
&{\Phi _p}(x) = \frac{{A\left[ {\Theta (b) - \Theta (b - x)} \right]}}{{2\alpha }}\left\{ {k_2^3\sin \left[ {{k_1}\left( {x - b} \right)} \right] - k_1^3\sin \left[ {{k_2}\left( {x - b} \right)} \right]} \right\}\\
& - \frac{{A\left[ {\Theta (b) - \Theta (b + x)} \right]}}{{2\alpha }}\left\{ {k_2^3\sin \left[ {{k_1}\left( {x + b} \right)} \right] - k_1^3\sin \left[ {{k_2}\left( {x + b} \right)} \right]} \right\},\\
&{\Psi _p}(x) = \frac{{A\left[ {\Theta (b) - \Theta (b - x)} \right]}}{{2\alpha }}\left\{ {{k_1}\sin \left[ {{k_2}\left( {x - b} \right)} \right] - {k_2}\sin \left[ {{k_1}\left( {x - b} \right)} \right]} \right\}\\
& - \frac{{A\left[ {\Theta (b) - \Theta (b + x)} \right]}}{{2\alpha }}\left\{ {{k_1}\sin \left[ {{k_2}\left( {x + b} \right)} \right] - {k_2}\sin \left[ {{k_1}\left( {x + b} \right)} \right]} \right\}.
\end{align}
\end{subequations}

The particular solution for Dirac delta charge may be further generalized to the 1D Kronig-Penney model which finds numerous applications in solid state physics. The particular solutions for the case of Dirac delta Kronig-Penney distribution ($b>0$) is
\begin{subequations}\label{soldkp}
\begin{align}
&{\Phi _p}(x) = \sum\limits_{m =  1}^{ + \infty } {\frac{{A\left[ {\Theta (mb) - \Theta (mb - x)} \right]}}{{2\alpha }}\left\{ {k_2^3\sin \left[ {{k_1}\left( {x - mb} \right)} \right] - k_1^3\sin \left[ {{k_2}\left( {x - mb} \right)} \right]} \right\}}\\
&- \sum\limits_{m =  0 }^{ + \infty }\frac{{A\left[ {\Theta (mb) - \Theta (mb + x)} \right]}}{{2\alpha }}\left\{ {k_2^3\sin \left[ {{k_1}\left( {x + mb} \right)} \right] - k_1^3\sin \left[ {{k_2}\left( {x + mb} \right)} \right]} \right\},\\
&{\Psi _p}(x) = \sum\limits_{m =  1}^{ + \infty } {\frac{{A\left[ {\Theta (mb) - \Theta (mb - x)} \right]}}{{2\alpha }}\left\{ {{k_1}\sin \left[ {{k_2}\left( {x - mb} \right)} \right] - {k_2}\sin \left[ {{k_1}\left( {x - mb} \right)} \right]} \right\}}\\
&- \sum\limits_{m =  0 }^{ + \infty }\frac{{A\left[ {\Theta (mb) - \Theta (mb + x)} \right]}}{{2\alpha }}\left\{ {{k_1}\sin \left[ {{k_2}\left( {x + mb} \right)} \right] - {k_2}\sin \left[ {{k_1}\left( {x + mb} \right)} \right]} \right\}.
\end{align}
\end{subequations}
The initial values $\Phi_0$ and $\Psi_0$ in the general solution should also be set in a way that the $G'=2\pi/b$ is the reciprocal lattice vector of 1D Kronig-Penney plasmonic crystal, i.e.,
\begin{subequations}\label{kpg}
\begin{align}
&{\Phi _g}(x) = - \frac{{B\left( {{G'^2} - k_1^2 - k_2^2} \right)\cos \left( {G'x} \right)}}{{\left( {{G'^2} - k_1^2} \right)\left( {{G'^2} - k_2^2} \right)}},\\
&{\Psi _g}(x) = - \frac{B\cos \left( {G'x} \right)}{{\left( {{G'^2} - k_1^2} \right)\left( {{G'^2} - k_2^2} \right)}}.
\end{align}
\end{subequations}

Figure 2 shows the plasmon excitations with given energy in the presence of a Dirac delta sheet at the origin. Figure 2(a) depicts the plasmon excitation in the presence of an extremely narrow negatively charged sheet at the origin. The thin curve shows the plasmon excitation with the same energy in the absence of the charge sheet. It is clearly remarked that the charge sheet significantly modifies the plasmon electrostatic field variations in the whole space around the sheet. However, the presence of Dirac charge sheet leads to a discontinuity of electric field at the place of sheet. Moreover, Fig. 2(b) depicts the plasmon excitation in space around a positively charged sheet. It is clearly remarked that the plasmon electrostatic potential around the positive sheet is relatively higher compared to the previous case with negatively charged sheet. On the other hand, Figure 2(c) and 2(d) depict the plasmon wavefunction around the same sheets represented in Figs. 2(a) and 2(b), respectively. For the case of positive/negative charges the profile seems to be mirror reflected in places far from the origin where the sheet resides. However, the inspection of the values of $\Psi^2(x)$ around the charge sheet reveals that electrons are more localized around the positive sheet rather than that of negative charge.

The quantum plasmon excitation profiles in driven pseudoforce model and in the presence of double Dirac delta sheet is depicted in Fig. 3. In Fig. 3(a) and 3(b) It is seen that the wavefunction and electrostatic potential amplitudes inside the parallel charged sheets is comparatively lower than those outside this region. It is remarked that the electrostatic potential contributed from the charged sheets is only present outside of the sheet region. In other words the plasmon excitations inside the parallel charged sheets is isolated from the outside fields. This feature is found to be dominant even with the asymmetric charged sheet configurations (not shown here). Figure 3(c) shows the particular solution due to the charged sheets only which reveals that this solution does not contribute to the plasmon excitations inside the sheet region. Moreover, the plasmon excitations around the Dirac charged sheets for beating energy eigenvalue $E\simeq 1$ is depicted in Fig. 3(d) for both wavefunction and electrostatic potential profiles. It is remarked that the presence of charged sheets leads to amplification of beating phenomenon outside of the sheet region.

Figure 4 depict the effect of Dirac charge sheet on one dimensional plasmon excitations in a sinusoidal periodic lattice of spacing $a$. Figure 4(a) shows the plasmon excitation with the energy $E=5$ in unit of the plasmon energy $E_p=\hbar\omega_p$ where $\omega_p$ is the electron plasma oscillation frequency, around a negative charge sheet at the origin. The thin curve represents the period field of the lattice. It is remarked that driven plasmon excitations carries the periodicity of the lattice modulated over a large amplitude due to the charge sheet. The current problem may represent a one dimensional impurity screening in a plasmonic crystal. Because of the one dimensional screening instead of realistic spherical problem the decaying does not take place. The plasmon excitation around a positive sheet with the same charge, lattice parameter and plasmon energy is seen in Fig. 4(b). The electrostatic potential around the sheet has become more positive in Fig. 4(b) due to the presence of positively charged sheet. Moreover, Fig. 4(c) shows the wavefunction profile quantum electron plasma oscillation in a plasmonic crystal around the negatively charged Dirac sheet at the origin with same parameters as in Fig. 4(a). It is remarked that, the electronic density is localized at the ionic positions. However, it is shown that the wavefunction and electrostatic potential are out of phase everywhere in plasmonic lattice. In Fig. 4(d), on the other hand, the negative charge sheet has been replaced with a positive sheet of same charge but with identical other parameters. It is remarked that the free electronic charge in this case as compared to Fig. 4(c) is more localized around the positive sheet.

The time evolution of the plasmon excitation in lattice containing dirac charge sheet appears in Fig. 5. Figure 5(a) shows perfect periodic plasmon excitations in the absence of charge distributions. As it is clear the Fig. 5(a) reflect a perfect harmonic pattern due to the lattice periodicity. The introduction of a weak negatively charged Dirac sheet at the origin, however, makes the clear periodic pattern somehow blurry but still the crystal periodicity is apparent from Fig. 5(b). With further increase in the charge of the Dirac sheet, on the other hand, the periodicity is further destroyed in Fig. 5(c). Figure 5(d) shows the time evolution of plasmon excitation with elevated energy level as compared to Fig. 5(d). It is remarked that plasmon excitations with higher energy values feel less that those of lower energy the lattice periodicity in the presence of a Dirac charge sheet.

\section{Conclusion}

Using the driven coupled pseudoforce system derived from one dimensional Schr\"{o}dinger-Poisson model the plasmon excitations of a free electron gas with arbitrary degree of degeneracy and electron fluid temperature is studied in the presence of various charge distributions. Closed forms of analytic functions is developed for plasmon excitations in the presence of an arbitrary charge distribution. The developed model is shown to be completely flexible to model a variety of physical problems with wide range of applications such as the impurity screening in 1D lattice, multi-charge distributed layers, etc. Current research help to study in detail the nature of quantum free electron excitations in periodic and quasi-periodic charged environments like plasmonic lattice and in presence of arbitrary and random impurity distributions. The model can also be further extended to include complex charge distributions containing arbitrary shaped charge regions fitting the desired physical situations.

\section{Acknowledgement}

Some analytic calculations in this research has been performed and checked for consistency by the Wolfram Mathematica symbolic computing software \cite{mathematica}.

\end{document}